\documentclass[10pts,showpacs,preprint,aps]{revtex4}
\usepackage{graphicx}
\usepackage{dcolumn}
\usepackage{bm}
\usepackage{amssymb}
\usepackage{amsmath}
\begin{document}
\setcounter{page}{1}
\vskip 2cm
\title
{On the Black Holes in alternative theories of gravity: The case of non-linear massive gravity}
\author
{Ivan Arraut$^{(1,2)}$}
\affiliation{$^1$Department of Physics, Osaka University, Toyonaka, Osaka 560-0043, Japan}
\affiliation{$^2$Theory Center, Institute of Particle and Nuclear Studies, KEK Tsukuba, Ibaraki, 305-0801, Japan}

\begin{abstract}
I derive general conditions in order to explain the origin of the Vainshtein radius inside dRGT. The set of equations, which I have called "Vainshtein" conditions are extremal conditions of the dynamical metric ($g_{\mu\nu}$) containing all the degrees of freedom of the theory. The Vainshtein conditions are able to explain the coincidence between the Vainshtein radius in dRGT and the scale $r_0=\left(\frac{3}{2}r_s r_\Lambda^2\right)^{1/3}$, obtained naturally from the Schwarzschild de-Sitter (S-dS) space inside General Relativity (GR). In GR, this scale was interpreted as the maximum distance in order to get bound orbits. The same scale corresponds to the static observer position if we want to define the black hole temperature in an asymptotically de-Sitter space. In dRGT, the scale marks a limit after which the extra degrees of freedom of the theory become relevant.
\end{abstract}
\maketitle 
\section{Introduction}
The Schwarzschild de-Sitter (S-dS) space in static coordinates has been widely studied in the past. Its analytic extension for S-dS space has been performed by Ba$\dot{z}a\acute{n}$ski and Ferrari \cite{1}. They interpreted the scale $r_0=\left(\frac{3}{2}r_s r_\Lambda\right)^{1/3}$ as the distance where the 0-0 component of the S-dS metric takes a minimum value. As a consequence of this, it was found in \cite{2} that $r_0$ represents a transition distance after which a photon suffers a gravitational blue shift when it moves away from a source.  
The same scale is used by Bousso and Hawking in order to find the appropriate expression for the temperature of a black hole immersed inside a de-Sitter space \cite{3}. In such a case, the distance $r_0$ is interpreted as the position of the static observer in order to find the appropriate normalization for the time-like Killing vector. Then there exist a minimum temperature for the black hole given by $T=\frac{1}{2\pi r_\Lambda}$ \cite{3,4}. This analysis differs in some details from the one done in \cite{S.W} where the Black Hole thermodynamics inside the S-dS space was analyzed in detail, in that case however, $r_0$ does not play the central role for the definition of the Black Hole temperature. The role of $r_0$ as a static radius was also analyzed in \cite{PRDsa2} inside the Kerr-de Sitter space.
In \cite{5}, Balaguera et al, found that $r_0$ represents the maximum distance within which we can find bound orbits solutions for a test particle moving around a source. In the same manuscript, the velocity bounds for a test particle inside the S-dS space were obtained, this work was then extended by Arraut et al in \cite{6} in order to incorporate other metric solutions. In \cite{5}, the authors also found that there exist a maximum angular momentum $L_{max}$ for the test particle to be inside a bound orbit. If $L=L_{max}$, then there exist a saddle point for the effective potential at the distance $r_x$, this analysis was extended recently by the author \cite{Mine}. In \cite{Bull1} and \cite{PRDsa}, the scale $r_0$ was derived by using a different method and some conditions for the circular orbits and its stability conditions were obtained. However, in such a case, the conditions were not interpreted in terms of a maximum angular momentum $L_{max}$. 
The scale $r_0$ plays a central role inside the $\Lambda_3$ version of the non-linear theory of massive gravity where it represents the distance below which non-linearities become important and General Relativity is restored \cite{deRham}. In this paper I derive general conditions in order to explain the origin of the Vainshtein radius and its coincidence with the same scale obtained inside the General Relativity (GR) formulation. I have called them the "Vainshtein" conditions, which correspond to extremal conditions for the components of the dynamical metric when all the degrees of freedom are inside it. 
The justification of this result is related to the fact that the massive potential $(U(g,\phi))$ in dRGT is a polynomial contraction between the dynamical metric and the fiducial one $(f_{\mu\nu})$. Then the Vainshtein scale emerging as an extremal condition for the massive potential is equivalent to the same conditions, but expressed in terms of the dynamical metric with all the degrees of freedom (5 in total). This result is demonstrated inside this manuscript and provides the central point of the analysis. 
For completeness, I analyze the equations of motion for a massive test particle under the influence of the S-dS metric in dRGT. I find that the equations of motion contain a velocity-dependent effective potential, suggesting then that the total energy is not conserved in its usual form. However, the notion of energy can be extended ($E_{dRGT}$) and then the equations of motion can be written in terms of this variable. If this is done, the resulting equations will not differ in essence with respect to the equations found in the standard Einstein theory of gravity, at least at the background level. 
The paper is organized as follows: In Section (\ref{eq:dRGT}), I introduce the basic aspects of the S-dS space in static coordinates and I then derive the scale $r_0$ including its correction due to the angular momentum of a massive test particle moving around the source. In Section (\ref{eq:2S}), I analyze the Black Hole temperature for an asymptotically de-Sitter space as has been defined by Bousso and Hawking. I explain the role of the scale $r_0$ in that situation. In Section (\ref{eq:dRGTTTTT}), I introduce the S-dS solution derived from the non-linear theory of massive gravity and then I explain the role of $r_0$ in this theory. In section (\ref{eq:Final1}), I write the S-dS solution inside dRGT gravity as has been derived by Kodama and the author, the solution written in this form is the most generic one for the spherically symmetric situations. In section (\ref{eq:Final2}) I derive the Vainshtein conditions in order to explain why $r_0$ appears in both formulations, namely GR and dRGT even if both theories are in principle different. Although the results are obtained for the S-dS solution in dRGT, they can be applied to any other solution in dRGT or any other theory of massive gravity. In section (\ref{eq:Lambda5}), I use the Vainshtein conditions in order to derive the Vainshtein scale inside the $\Lambda_5$ theory of massive gravity. In section (\ref{eq:JustVainshtein}), I demonstrate that the extremal condition of the dynamical metric with all the degrees of freedom inside, is just equivalent to an extremal condition of the massive action $(U(g,\phi))$, providing then an alternative way for the derivation of the Vainshtein scale. In such calculation it is enough to work with terms of quadratic order because the higher order contributions will contain exactly the same scale. In section (\ref{eq:Final33}), I derive the equations of motion of a massive test particle under the influence of the S-dS solution in dRGT. In such a case, it is demonstrated that the dynamics of a test particle moving around a spherically symmetric source is the same as in GR if we are able to extend the notion of energy in dRGT. Finally, in section (\ref{eq:conclusions}), I conclude. 
           
\section{The Schwarzschild de-Sitter space}   \label{eq:dRGT}

The Schwarzschild-de Sitter metric in static coordinates, is given by:

\begin{equation}   \label{eq:Sdsm}
ds^2=-e^{\nu(r)}dt^2+e^{-\nu(r)}dr^2+r^2d\theta^2+r^2\sin^2\theta d\phi^2,
\end{equation}

where:

\begin{equation}   \label{eq:e}
e^{\nu(r)}=1-\frac{r_s}{r}-\frac{r^2}{3r_\varLambda^2},
\end{equation}

where $r_s=2GM$ is the gravitational radius and $r_\Lambda=\frac{1}{\sqrt{\Lambda}}$ defines the cosmological constant scale. In this coordinate system, it has been demonstrated that the effective potential is given by:

\begin{equation}   \label{eq:effpotaa}
U_{eff}(r)=-\frac{r_s}{2r}-\frac{1}{6}\frac{r^2}{r_\varLambda^2}+\frac{L^2}{2r^2}-\frac{r_sL^2}{2r^3},
\end{equation}

where $U_{eff}(r)$ is the effective potential which influences the motion of a massive test particle in S-dS space. The equation of motion of a massive test particle is given by:

\begin{equation}   \label{eq:Moveq2}
\frac{1}{2}\left(\frac{dr}{d\tau}\right)^2+U_{eff}(r)=\frac{1}{2}\left(E^2+\frac{L^2}{3r_\varLambda^2}-1\right)=C,
\end{equation}

where $C$ is a constant depending on the initial conditions of motion. This effective potential has three circular orbits. They correspond to the condition $\frac{dU_{eff}(r)}{dr}=0$. In this manuscript, I focus on the scale $r_0$ which corresponds to one of the previously mentioned circular orbits. In \cite{Mine}, $r_0$ was derived and it is given by: 

\begin{equation}   \label{eq:srwbimbto}
r_0(\beta)=\left(\frac{3}{2}r_sr_\varLambda^2\right)^{1/3}-\frac{1}{4\beta^2}(3r_sr_\varLambda^2)^{1/3},
\end{equation}

where we make explicit the angular momentum dependence of the massive test particle through the parameter $\beta=L/L_{max}$, with $L_{max}=\frac{{3}^{2/3}}{4}(r_s^2r_{\varLambda})^{1/3}$ being the maximum angular momentum if we want to get bound orbits. This scale is the limit where the attractive effects due to gravity and the repulsive ones due to the cosmological constant ($\Lambda$) just cancel. This is the key point in the Bousso-Hawking definition of temperature as will be explained in the next section.

\section{Black Hole thermodynamics in an asymptotically de-Sitter space: The role of the scale $r_0$}   \label{eq:2S}

In agreement with Bousso and Hawking, the appropriate way to define the black hole thermodynamics is by normalizing the time-like Killing vector such that the static observer is located at the distance given by (\ref{eq:srwbimbto}) with $\beta=0$. If we assume that the observer does not have any angular momentum, then the surface gravity is defined as \cite{3}:

\begin{equation}   \label{eq:12}
\kappa_{BH, CH}=\left(\frac{(K^\mu\nabla_\mu K_\gamma)(K^\alpha\nabla_\alpha K^\gamma)}{-K^2}\right)^{1/2}_{r=r_{BH},r_{CH}}.
\end{equation}

The subindices BH and CH, correspond to the Black Hole Horizon and the Cosmological one respectively. The event horizons are obtained from the condition:

\begin{equation}   \label{eq:51}
g^{rr}(r_c)=0,
\end{equation} 

and they are given explicitly by:

\begin{equation*}   
r_{CH}=-2r_\varLambda cos\left(\frac{1}{3}\left(cos^{-1}\left(\frac{3r_s}{2r_\varLambda}\right)+2\pi\right)\right),
\end{equation*}

\begin{equation*}
r_{BH}=-2r_\varLambda cos\left(\frac{1}{3}\left(cos^{-1}\left(\frac{3r_s}{2r_\varLambda}\right)+4\pi\right)\right).
\end{equation*} 

The two horizons become equal when the mass of the Black Hole reach its maximum value given by:

\begin{equation}   \label{eq:8}
M_{max}=\frac{1}{3}\frac{m_{pl}^2}{m_\varLambda},
\end{equation}

where $m_{pl}$ corresponds to the Planck mass and $m_\Lambda=\sqrt{\Lambda}$. If the mass of a Black Hole is larger than the value given by eq. (\ref{eq:8}), then there is no radiation at all and we have a naked singularity. As $M=M_{max}$, the two event horizons take the same value $\left(r_{BH}=r_{CH}=r_\Lambda=\frac{1}{\sqrt{\Lambda}}\right)$, they are degenerate and a thermodynamic equilibrium is established.
As has been explained by Bousso and Hawking \cite{3}, as $M\to M_{max}$, $V(r)\to0$ between the two horizons (BH and Cosmological) and for that reason the Schwarzschild-like coordinates simply become inappropriate. In such a case we need a new coordinate system. In agreement with Ginsparg and Perry \cite{GP}, we write:

\begin{equation}   \label{eq:9}
9M^2\Lambda=1-3\epsilon^2,   \;\;\;\;\;0\leq\epsilon\ll1,
\end{equation} 

where $\epsilon$ is a parameter related to the mass of the black-hole. In these coordinates, the degenerate case (when the two horizons become the same), corresponds to $\epsilon\to0$. We must then define the new radial and the new time coordinates to be:

\begin{equation}   \label{eq:10}
\tau=\frac{1}{\epsilon\sqrt{\Lambda}}\psi,\;\;\;\;\; r=\frac{1}{\sqrt{\Lambda}}\left(1-\epsilon cos\chi-\frac{1}{6}\epsilon^2\right).
\end{equation}

In these coordinates, the Black Hole horizon corresponds to $\chi=0$ and the Cosmological horizon to $\chi=\pi$ \cite{3}. The new metric obtained from the transformation is given by:

\begin{equation}   \label{eq:11}
ds^2=-r_\Lambda^2\left(1+\frac{2}{3}\epsilon cos \chi\right)sin^2\chi d\psi^2+r_\Lambda^2\left(1-\frac{2}{3}\epsilon cos\chi\right)d\chi^2+r_\Lambda^2(1-2\epsilon cos\chi)d\Omega_2^2.
\end{equation}  

This metric has been expanded up to first order in $\epsilon$. Eq. (\ref{eq:11}) is of course the appropriate metric to be used as the mass of the Black Hole is near to its maximum value given by eq. (\ref{eq:8}). It has been found by Bousso and Hawking that the time-like Killing vector inside the definition (\ref{eq:12}) has to be normalized in agreement with:

\begin{equation}   \label{eq:16}
\gamma_t=\left(1-\left(\frac{3r_s}{2r_\Lambda}\right)^{2/3}\right)^{-1/2},
\end{equation}

with $\gamma_t$ being the normalization factor for the time-like Killing vector defined as:

\begin{equation}   \label{eq:13}
K=\gamma_t\frac{\partial}{\partial t}.
\end{equation}

In an asymptotically flat space, $\gamma_t\to1$ when $r\to\infty$. But in the case of eq. (\ref{eq:16}), the Killing vector is just normalized with respect to an observer at the position $r_0$ with $\beta=0$ as has been defined previously. When the mass of the black hole reach its maximum value defined as $\epsilon\to0$ in eq. (\ref{eq:9}), the black hole temperature reach its minimum value given by:

\begin{equation}   \label{eq:19}
2\pi T_{min=}\kappa_{min}^{BH}=\frac{1}{r_\Lambda},
\end{equation}   

where $\kappa$ is the surface gravity. Note the importance of the scale $r_0$ in the definition of the black-hole temperature in this case.  

\section{Black Holes in dRGT non-linear theory of massive gravity}   \label{eq:dRGTTTTT}

In agreement with Koyama and colleagues, it is possible to construct black hole solutions inside the non-linear theory of massive gravity. It is natural to suspect that such solution should be related in some sense to the S-dS solution of GR. However, other solutions are possible in principle. In dRGT, we can get the same solution given by eq. (\ref{eq:Sdsm}) but surrounded by a halo of helicity 0 and $\pm1$. The trick is to use as a starting point a metric of the form \cite{Gaba}:

\begin{equation}   \label{eq:20}
ds^2=-dt^2+(dr\pm\sqrt{f(r)}dt)^2+r^2d\Omega^2,
\end{equation}
 
where $f(r)$ will be defined later. The previous metric is free of horizon singularities, such that the invariant $g^{\mu\nu}\partial_\mu\phi^a\partial_\nu\phi^b\eta_{ab}$ (defined inside the dRGT theory) remains finite when all the other standard relativistic invariants are also finite. The metric has to be a solution of the Einstein's equations, which in massive gravity are defined as:

\begin{equation}   \label{eq:5}
G^\mu_{\;\;\nu}=-m^2X^\mu_{\;\; \nu}.
\end{equation}

The solution (\ref{eq:20}), after the appropriate coordinate transformations, becomes the same solution given by eq. (\ref{eq:Sdsm}) but surrounded by a St\"uckelberg (gravitational) background defined by:

\begin{eqnarray}   \label{eq:pp}
\Phi^0=\frac{1}{\kappa}(t+f(r)),\nonumber&&\\
\Phi^r=\left(1+\frac{1}{\alpha}\right)r,\nonumber&&\\
\Phi^\theta=\theta\nonumber,&&\\
\Phi^\phi=\phi.
\end{eqnarray}

The previous results correspond to a family of solutions satisfying a specific relation between the two free parameters of the theory as has been explained in \cite{Gaba, Gaba2}. The scale $r_0$ defined before, inside the $\Lambda_3$ version of the theory, appears as the Vainshtein radius if we tune the mass of the graviton with the $\Lambda$ scale. For distances satisfying the condition $r<<r_0$, non-linearities become relevant and General Relativity is recovered, avoiding in such a way the vDVZ discontinuity \cite{Gaba3}. The non-linear solution inside the dRGT theory, admits perturbative expansions in terms of the mass of the graviton for distances satisfying $r<<r_0$. On the other hand, the same solutions admit perturbative expansions in terms of the Newtonian Constant for distances $r>>r_0$. Then in some sense, $r_0$ is a scale which marks the transition between a solution dominated by the Newtonian constant and the one dominated by the graviton mass in direct analogy with what happens in General Relativity when we compare the Newtonian constant with the Cosmological Constant scale $(\Lambda)$. The main difference is that $r_0$ in massive gravity is related to the existence of a strong coupling scale $\Lambda_3=(M_{pl}m^2)^{1/3}$ which appears in the Lagrangian when the theory is ghost-free \cite{Gaba2}. In fact, when the dRGT theory was discovered, the coefficients of the massive potential were tuned such that the theory became ghost-free \cite{Ghostsol}. Although the authors of the dRGT formulation of massive gravity did a remarkable job, the same result can be obtained by reasoning in a different way. If for example we define some scale of duality defined by the extreme condition of the massive potential $(U(g,\phi))$, then in principle it would be possible to tune the coefficients of the potential, such that the duality scale, given by the Vainshtein radius, takes the value given by the combination $r_V=(GM/m^2)^{1/3}$. When this combination is satisfied, then the theory becomes automatically ghost-free. The purpose of this manuscript is then, to provide a deep understanding of the scale $r_V$, which is a manifestation of the duality between the scales given by the Newtonian constant and the mass of the graviton respectively. 

\section{The Schwarzschild de-Sitter solution in dRGT} \label{eq:Final1}

In \cite{Kodama}, the S-dS solution was derived for two different cases. The first one, corresponds to the family of solutions satisfying the condition $\beta=\alpha^2$, where $\beta$ and $\alpha$ correspond to the two free-parameters of the theory. In such a case, the "gauge" transformation function $T_0(r,t)$ becomes arbitrary. The second one, corresponds to the family of solutions with two-free parameters satisfying the condition $\beta\leq\alpha^2$ with the "gauge" transformation function $T_0(r,t)$ constrained. The generic solution is given explicitly as:

\begin{equation}
ds^2=g_{tt}dt^2+g_{rr}dr^2+g_{rt}(drdt+dtdr)+r^2d\Omega_2^2,
\end{equation}
 
where:

\begin{equation}   \label{eq:drgt metric}
g_{tt}=-f(r)(\partial_tT_0(r,t))^2,\;\;\;\;\;g_{rr}=-f(r)(\partial_rT_0(r,t))^2+\frac{1}{f(r)},\;\;\;\;\;g_{tr}=-f(r)\partial_tT_0(r,t)\partial_rT_0(r,t),
\end{equation}

where $f(r)=1-\frac{2GM}{r}-\frac{1}{3}\Lambda r^2$. In this previous solution, all the degrees of freedom are inside the dynamical metric. The fiducial metric in this case is just the Minkowskian one given explicitly as:

\begin{equation}
f_{\mu\nu}dx^\mu dx^\nu=-dt^2+\frac{dr^2}{S_0^2}+\frac{r^2}{S_0^2}(d\theta^2+r^2sin^2\theta),
\end{equation}
 
where $S_0=\frac{\alpha}{\alpha+1}$. The St\"uckleberg fields take the standard form defined in \cite{Kodama}. Note that the function $T_0(r,t)$
is not a gauge function in the sense of dRGT because it appears after the introduction of the extra-degrees of freedom by using the St\"uckelberg trick. In fact, the extra-degrees of freedom enter in a similar way as coordinate transformations from the GR point of view. Those "transformations" however, do not correspond to dipheomorphism transformations from the dRGT point of view. The $T_0(r,t)$ function appears as:

\begin{equation}
g_{\mu\nu}=\left(\frac{\partial Y^\alpha}{\partial x^\mu}\right)\left(\frac{\partial Y^\beta}{\partial x^\nu}\right)g'_{\alpha\beta},
\end{equation}  

with:

\begin{equation}
Y^0(r,t)=T_0(r,t),\;\;\;\;\;Y^r(r,t)=r.
\end{equation}

\section{The Vainshtein conditions}   \label{eq:Final2}

In S-dS space (in static coordinates), the scale $r_0$ with $\beta=0$ can be obtained by solving the equation obtained from the condition $df(r)/dr=0$. The solution shows the distance after which the slope of the function $f(r)$ changes its signature. This is not a coincidence since it is evident that the scale after which the cosmological constant becomes "dominant" has to be marked by an extremal condition. 
In the case where the metric components depend on both, position and time, the extremal conditions can be written in the following form:

\begin{equation}   \label{eq:cond1}
dg_{\mu\nu}=\left(\frac{\partial g_{\mu\nu}}{\partial r}\right)_t dr+\left(\frac{\partial g_{\mu\nu}}{\partial t}\right)_r dt=0,
\end{equation}       

where the notation $\left(\frac{\partial g_{\mu\nu}}{\partial w}\right)_x$ is just the partial derivative with respect to $w$ but keeping the variable $x$ constant. When applied to a static metric solution, eq. (\ref{eq:cond1}) is equivalent to $\partial g_{\mu\nu}/\partial r=0$. In the case of S-dS space in GR, the condition (\ref{eq:cond1}) is satisfied as $r=r_0$ with $\beta=0$. 
We can apply the same extremal conditions to the dynamical metric solutions of dRGT theory when all the degrees of freedom (5 in total) are contained inside it. In dRGT, the extremal condition (\ref{eq:cond1}) marks the scale after which the metric behavior begins to change and as a consequence, it marks the scale after which the extra-degrees of freedom become relevant. This scale is in fact the Vainshtein scale (Vainshtein radius for static metrics). If we introduce the metric defined in (\ref{eq:drgt metric}) inside the condition (\ref{eq:cond1}), then we obtain the Vainshtein conditions for the metric under study. For the $t-t$ component, we get:

\begin{equation}   \label{eq:verg}
\left(f'(r)(\partial_tT_0(r,t))^2+2f(r)(\partial_tT_0(r,t))(\partial_r\partial_tT_0(r,t))\right)dr+2f(r)(\partial_tT_0(r,t))\partial^2_tT_0(r,t)dt=0.
\end{equation}

For the quasi-stationary case (i.e, the case where the metric can be translated to static coordinates), $T_0(r,t)\backsim t+A(r)$, where $A(r)$ is an arbitrary function on space. Then the previous condition is reduced to $f'(r)=0$. For the $r-r$ component, we have to satisfy:

\begin{eqnarray}
\left(f'(r)(\partial_rT_0(r,t))^2+2f(r)(\partial_rT_0(r,t))(\partial^2_rT_0(r,t))+\frac{f'(r)}{f(r)^2}\right)dr\nonumber\\
+2f(r)(\partial_rT_0(r,t))(\partial_t\partial_rT_0(r,t))dt=0.
\end{eqnarray}

Again, if we assume quasi-stationary condition, then the previous result is reduced to:

\begin{equation}   \label{eq:thisoneag}
T_0'(r,t)\left(f'(r)T_0'(r,t)+2f(r)T_0''(r,t)\right)+\frac{f'(r)}{f(r)^2}=0.
\end{equation}

Finally, for the $t-r$ component, we get:

\begin{eqnarray}   \label{eq:solvef'}
(f'(r)(\partial_tT_0(r,t))(\partial_rT_0(r,t))+f(r)(\partial_t\partial_rT_0(r,t))(\partial_rT_0(r,t))+f(r)(\partial_tT_0(r,t))\times\nonumber\\
(\partial^2_rT_0(r,t)))dr+\left(f(r)(\partial^2_tT_0(r,t))(\partial_rT_0(r,t))+f(r)(\partial_tT_0(r,t))(\partial_r\partial_tT_0(r,t))\right)dt=0.
\end{eqnarray}

Once again, if the dynamical metric is quasi-stationary, then we get:

\begin{equation}   \label{eq:solvef2}
\frac{T''_0(r,t)}{T'_0(r,t)}=-\frac{f'(r)}{f(r)}=C,
\end{equation}

where $C$ is an arbitrary constant and we assume $\partial_t^2T_0(r,t)=\partial_t\partial_r T_0(r,t)=0$. Eq. (\ref{eq:solvef2}) can be solved by separation of variables. By assuming a general exponential behavior, we would get:

\begin{equation}   \label{eq:solvef3}
T_0(r,t)=Ae^{C r},\;\;\;\;\;f(r)=Be^{-C r}.
\end{equation}

If we replace the condition (\ref{eq:solvef2}) inside eq. (\ref{eq:thisoneag}), then we get:

\begin{equation}
T_0'(r,t)=\pm f^{-1},
\end{equation}

which is consistent with the result (\ref{eq:solvef3}). This also implies that under the quasi-stationary condition, at the Vainshtein radius, $T''_0(r,t)=0$ in agreement with the result obtained from (\ref{eq:verg}). The set of conditions condensed in the single expression (\ref{eq:cond1}) is what I have called "Vainshtein conditions". From them we can find the Vainshtein scale which can be time-dependent for general backgrounds. The Vainshtein scale becomes equivalent to the already known Vainshtein radius when the metric is time-independent. In such a case, the result (\ref{eq:cond1}) is reduced to $\partial g_{\mu\nu}/\partial r=0$, obtaining then as a Vainshtein radius the result $r_V=r_0$. This explains the origin of the coincidence with respect to the scale $r_0$ obtained from GR. Just for completing the previous arguments, we can specify that the Vainshtein scale is marked by three regimes:

\begin{eqnarray}   \label{eq:setofcond}
\partial_rT_0(r,t)=0\;\;\;\;\;\to\;\;\;\;\;r<<r_V,\;\;\;\;\;\partial_r T_0(r,t)\neq0\;\;\;\;\;\to\;\;\;\;\;r>>r_V\nonumber\\
T_0''(r,t)=0\;\;\;\to\;\;\;\;\;r=r_V,
\end{eqnarray}

where $r_V$ is just the Vainshtein scale.

\section{The Vainshtein radius in $\Lambda_5$ theory}   \label{eq:Lambda5}

It is easy to extend the concepts of the previous section to the $\Lambda_5$ theory of non-linear massive gravity. In such a case, it is known that the Vainshtein scale is different with respect to the the case of $\Lambda_3$ and as a consequence, $r_V$ will differ with respect to the case of GR with $\Lambda$. However, what is necessary to remark is that the origin of the Vainshtein scale again emerges as an extremal condition for the dynamical metric with all the degrees of freedom. If we write the metric in a diagonal form as \cite{Gaba3}:

\begin{equation}
ds^2=-B(r)dt^2+A(r)dr^2+r^2C(r)d\Omega^2,
\end{equation}

where the fiducial metric still keeps the Minkowskian form. In this case, it is also easy to verify that the conditions:

\begin{equation}
B'(r)=A'(r)=\frac{d(r^2C(r))}{dr}=0,
\end{equation}

reproduce the Vainshtein radius inside this theory. It is given by $r_V=(GM/m^4)^{1/5}$. In fact, it is possible to verify that we can reproduce the Vainshtein scale for any massive theory of gravity by using the same principles. If gravity disappears, then the Vainshtein scale vanishes, namely, $r_V=0$. This only means that the extra-degrees of freedom of the theory are relevant at any scale because there is no strong coupling scale able to reproduce non-linearities in order to screen the affects of the extra-degrees of freedom of the theory.  

\section{The justification of the Vainshtein conditions: Extreme conditions on the massive potential}   \label{eq:JustVainshtein}

In the previous two sections, I introduced the so-dubbed Vainshtein conditions. They correspond to extremal conditions on the dynamical metric when it contains all the degrees of freedom (5 in total) of the theory. When the dynamical metric contains all the degrees of freedom, it becomes dipheomorphism invariant by itself. In this section I want to justify why the Vainshtein scale appears as the extremal condition for the dynamical metric ($g_{\mu\nu}$) (omly when the dynamical metric has all the degrees of freedom). For that purpose I will concentrate on the massive action $U(g,\phi)$. In fact, the strong coupling scale will appear implicitly contained inside the massive action \cite{Gaba3, Ghostsol}. At the basic level, this means that there will be a scale below which the extra-degrees of freedom will become negligible. This quality of the theory is known as the Vainshtein mechanism and has been studied widely in the past \cite{deRham}. The novelty in this paper is that the mechanism is introduced as a set of conditions on the dynamical metric and as a consequence, as a set of conditions on the "gauge" function ($T_0(r,t)$). The modified Einstein-Hilbert action is given by:

\begin{equation}   \label{eq:b1}
S=\frac{1}{2\kappa^2}\int d^4x\sqrt{-g}(R+m^2U(g,\phi)),
\end{equation}

with the effective potential depending on two free parameters as:

\begin{equation}   \label{eq:b2}
U(g,\phi)=U_2+\alpha_3 U_3+\alpha_4U_4,
\end{equation}

where:

\begin{equation}   \label{eq:b3}
U_2=Q^2-Q_2,
\end{equation}

\begin{equation}   \label{eq:b4}
U_3=Q^3-3QQ_2+2Q_3,
\end{equation}

\begin{equation}   \label{eq:b5}
U_4=Q^4-6Q^2Q_2+8QQ_3+3Q_2^2-6Q_4,
\end{equation}

\begin{equation}   \label{eq:b6}
Q=Q_1,\;\;\;\;\;Q_n=Tr(Q^n)^\mu_{\;\;\nu},
\end{equation}

\begin{equation}   \label{eq:b7}
Q^\mu_{\;\;\nu}=\delta^\mu_{\;\;\nu}-M^\mu_{\;\;\nu},
\end{equation}

\begin{equation}   \label{eq:b8}
(M^2)^\mu_{\;\;\nu}=g^{\mu\alpha}f_{\alpha\nu},
\end{equation}

\begin{equation}   \label{eq:b9}
f_{\mu \nu}=\eta_{ab}\partial_\mu\phi^a\partial_\nu\phi^b.
\end{equation}    

In this case we can consider the fiducial metric as the Minkowskian one, due to the fact that all the degrees of freedom are contained inside the dynamical metric. The purpose here is to analyze the structure of $U(g,\phi)$. The potential (massive action) is in general given by \cite{Gaba3, Ghostsol}:

\begin{equation}   \label{eq:Qexpli}
U(g, \phi)=-4\left(<Q>^2-<Q^2>\right).
\end{equation}
   
From the definition (\ref{eq:b7}) and (\ref{eq:b8}), it is clear that an extremal condition for the potential $U(g, \phi)$, will correspond to an extremal condition for the dynamical metric ($g_{\mu\nu}$) if it contains all the degrees of freedom. Then:

\begin{equation}   \label{eq:ptm}
dU(g,\phi)=\left(\frac{\partial U(g, \phi)}{\partial g}\right)_\phi dg=0,
\end{equation}

is equivalent to:

\begin{equation}   \label{eq:this one}
dg=0,
\end{equation}

for all the components of the dynamical metric. The explicit evaluation of the total differential for the expression (\ref{eq:Qexpli}) by using (\ref{eq:b7}) and (\ref{eq:b8}) is not necessary because the extremal condition for of the root square of the matrix will be equivalent to the extremal condition for the matrix itself. In other words, the extremal condition for $\sqrt{g^{\mu\gamma}f_{\gamma\nu}}$ is the same as the extremal condition for $g^{\mu\gamma}f_{\gamma\nu}$. Then the apparently complicated (explicit) expression given by eq. (\ref{eq:ptm}) is reduced to the simple condition given by eq. (\ref{eq:this one}). The Vainshtein scale will emerge then as an extremal condition of the dynamical metric, if and only if, the dynamical metric contains all the degrees of freedom. This condition is necessary in order to guarantee that no degree of freedom is contained inside the fiducial metric which is just Minkowski. The mechanism formulated in this form, deserves more attention. 

\subsection{The massive action expansion} 

The massive action (\ref{eq:Qexpli}), when expanded, has to be equivalent to the action (\ref{eq:b2}) with the appropriate definitions. We can write $U(g,\phi)$ in terms of the $Q$-matrices defined previously, or equivalently, we can expand $U(g,\phi)$ in terms of a covariant object $H_{\mu\nu}$ defined as \cite{Gaba3, Ghostsol}:

\begin{equation}   \label{eq:Defini}
H_{\mu\nu}=\frac{h_{\mu\nu}}{M_{pl}}+\partial_\mu \pi_\nu+\partial_\nu \pi_\mu-\eta_{\alpha\beta}\partial_\mu \pi^\alpha\partial_\nu \pi^\beta,
\end{equation}

with the St\"uckelberg fields defined as:

\begin{equation}
\phi^a=x^a-\pi^a.
\end{equation}

This previous expansion is justified when the extra-degrees of freedom are stored inside the fiducial metric. In terms of the tensor $H_{\mu\nu}$, the $Q$-matrices are defined as:

\begin{equation}
Q^\mu_{\;\;\nu}=\delta^\mu_{\;\;\nu}-\sqrt{\delta^\mu_{\;\;\nu}-H^\mu_{\;\;\nu}}, 
\end{equation} 

and after expanding the root square and then introducing the results inside eq. (\ref{eq:Qexpli}) for the corresponding expansion of the massive action, then we get:

\begin{equation}
U(g, H)=-4\left(\sum_{n\geq1}\bar{d}_n<H^n>\right)^2-8\sum_{n\geq2}\bar{d}_n<H^n>,
\end{equation}

which is ghost-free. With the definition (\ref{eq:Defini}) inside this massive action and then using the re-definition $\pi^a=\partial^a\pi$, then we get the action with the Galileon structure as has been analyzed previously in \cite{Ghostsol}. For simplicity, here I write the simplest scalar action with the corresponding strong-coupling and given by:

\begin{equation}   \label{eq:Pounds}
\pounds=-\frac{1}{2}(\partial\pi)^2-\frac{1}{\Lambda^3}(\partial\pi)^2\square\pi+\frac{1}{M_{pl}}\pi T.
\end{equation}  

If we want to compute the Vainshtein scale $r_*$, all what we have to do is to compare the linear term with the non-linear one and after, it is necessary to make a final comparison with the source term. In other words, the Vainshtein scale corresponds to the scale where the three terms of the action (\ref{eq:Pounds}) have the same order of magnitude. For comparing the first two terms of the action which correspond to the linear and non-linear contribution, we have to calculate the extremal condition by calculating the total differential. In other words, let's define:

\begin{equation}   \label{eq:action this}
\pounds_{Vacuum}=-\frac{1}{2}(\partial\pi)^2-\frac{1}{\Lambda^3}(\partial\pi)^2\square\pi.
\end{equation}   

By assuming spherical symmetry and time-independence of the field $\pi(r)$, we can find the scale at which the action (\ref{eq:action this}) is extremal. The condition to be satisfied by the action (\ref{eq:action this}) is:

\begin{equation}
-(\pi')(\pi'')-\frac{2}{\Lambda^3}(\pi')(\pi'')^2-\frac{1}{\Lambda^3}(\pi')^2\pi'''=0,
\end{equation}

with the results:

\begin{equation}   \label{eq:results}
\pi'(r)=-\frac{\Lambda^3}{2}r_*,\;\;\;\;\;\pi''(r)=-\frac{\Lambda^3}{2}, \;\;\;\;\;\pi'''(r)=0,
\end{equation}

with $r_*$ defining the Vainshtein scale. If we replace this previous result in any of the first two terms of the action (\ref{eq:Pounds}), then we obtain a contribution with order of magnitude $\Lambda^6r_*^2$. We can then compare the result with the source contribution defined by:

\begin{equation}
T=-\frac{M}{4\pi r^2}\delta(r),
\end{equation}

but the delta function $(\delta(r))$ is peaked at the origin of the spherically symmetric source and not necessarily at the Vainshtein scale. Then in principle its contribution vanishes at the Vainshtein radius $r_*$. However, it is still possible to compare the source term with the order of magnitude $\Lambda^6r_*^2$ if we integrate over the whole spatial volume. For the comparison we can select the first term in the action (\ref{eq:Pounds}) and then use the source term. It is necessary to use the results (\ref{eq:results}) in order to guarantee that the linear and the non-linear contribution (at the vacuum level) are equivalent. After integration over the whole spatial volume and then sending the result to zero, we get:

\begin{equation}
-\frac{\Lambda^6}{16}r_*^3+\frac{M}{M_{pl}\pi}=0.    
\end{equation}

If we solve this previous equation, then we get:

\begin{equation}
r_*=\frac{1}{\Lambda}\left(\frac{M}{M_{pl}}\right)=\left(\frac{M}{\pi M_{pl}^2m^2}\right)^{1/3},
\end{equation}

where we have used the strong coupling definition $\Lambda^3=m^2M_{pl}$. Then the Vainshtein radius is in reality the scale at which all the contributions in the action are comparable. Then defining the Vainshtein scale as the scale at which the the dynamical metric is extremal, just simplifies the calculations. Note that the same result would be obtained if just differentiate the components of the dynamical metric. Physically this is a consequence of the change of geometry produced by the source and the presence of the extra-degrees of freedom in the dynamical metric. When the comparison is done by using the extremal condition of the dynamical metric, then it is easy because the source term appears with the dependence $M/r$ and not as a delta-function. In such a case, it is not necessary to separate the different contributions of the action in order to perform the comparison.  

\section{The effective potential in dRGT massive gravity}   \label{eq:Final33}

In order to compare massive gravity with General Relativity when a test particle moves around a source, it is important to derive the equations of motion for a massive test particle when it moves around a spherically symmetric source. The equations of motion can be written as:

\begin{equation}   \label{eq:Miau}
\frac{1}{2}\left(\frac{dr}{d\tau}\right)^2-E\left(\frac{g_{tr}}{g_{rr}g_{tt}}\right)\left(\frac{dr}{d\tau}\right)+\frac{L^2}{2r^2g_{rr}}=-\frac{1}{2g_{rr}}\left(\frac{E^2}{g_{tt}}+1\right),
\end{equation}

where $g_{tt}$ and $g_{rr}$ are defined in eq. (\ref{eq:drgt metric}). Note that as $\partial_rT_0(r,t)=0$, the previous equation is reduced to the result (\ref{eq:Moveq2}) if we use the metric given by (\ref{eq:drgt metric}). If we replace the metric components (\ref{eq:drgt metric}) inside (\ref{eq:Miau}), then we get:

\begin{eqnarray}   \label{eq:Miau3}
\frac{1}{2}\left(\frac{dr}{d\tau}\right)^2+\frac{\partial_rT_0(r,t)f(r)E}{\partial_tT_0(r,t)(f(r)^2(\partial_rT_0(r,t))^2-1)}\left(\frac{dr}{d\tau}\right)-\frac{L^2}{2r^2}\left(\frac{f(r)}{f(r)^2(\partial_rT_0(r,t))^2-1}\right)\nonumber\\
=\frac{1}{2(\partial_tT_0(r,t))^2(f(r)^2(\partial_rT_0(r,t))^2-1)}\left(f(r)(\partial_tT_0(r,t)^2)-E^2\right).
\end{eqnarray}

In eq. (\ref{eq:Miau}), the energy and angular momentum have been introduced in the usual sense in agreement with the results of Sec. (\ref{eq:dRGT}). The presence of a velocity dependent quantity in eq. (\ref{eq:Miau3}) shows that the effective potential which influences the motion of a test particle, is velocity-dependent. This dependence cannot be gauged away as in GR. Then the origin of the velocity term inside the effective potential, comes from the extra-degrees of freedom. The dependence of the effective potential with the velocity suggests that the total energy of a test particle is not conserved in the usual form. However, it is possible to formulate the equations of motion in terms of an extended notion of energy which is conserved. The new definition of total energy can be formulated in agreement with:

\begin{equation}   \label{eq:Ronaldo}
g_{tt}\left(\frac{dt}{d\tau}\right)+g_{rt}\left(\frac{dr}{d\tau}\right)=E_{dRGT},
\end{equation}

which is in agreement with the direction of the new Killing vector. In terms of this definition, the equation of motion given in eq. (\ref{eq:Miau3}), just becomes the same as in GR. Then we can conclude that as far as the three-dimensional motion is involved, then the equations of motion will not differ from those obtained in GR. However, as soon as some dynamics is involved in the system, for example, Hawking radiation for black-holes, perturbations of the metric, etc, then the effects of the extra-degrees of freedom will affect the motion of test particles around a the source.  

\section{Conclusions}   \label{eq:conclusions}
I derived the general conditions in order to obtain the Vainshtein scale inside the dRGT formulation of massive gravity. The Vainshtein scale is just an extremal condition for the dynamical metric with all the degrees of freedom of the theory. This explains the coincidence between the Vainshtein radius obtained from dRGT and the scale $r_0$ obtained in the standard GR theory for the S-dS solution. In fact, the scale $r_0$ in GR with $\Lambda\neq0$ appears as an extremal condition of the metric. Then the apparent coincidence between the scales $r_0$ in GR and $r_V$ in dRGT massive gravity is explained by the so-dubbed "Vainshtein" conditions discovered by the author in this manuscript. The same scale appears at the local physics level in other theories. In $f(R)$-gravity theories for example, this scale can appear for the cases when the condition $f(R)\approx R$ is satisfied at the weak-field limit approximation for constant Ricci scalar solutions \cite{suggested1}. This is true if the theory reproduces in this limit a de-Sitter like-behavior as it is the case in \cite{suggested2}, where $f(R)$-gravity theories are able to reproduce the Dark Energy effects without cosmological constant \cite{suggested3}. Then the scale $r_0$ will appear in terms of the parameters of the theory. For the example illustrated in \cite{suggested2}, for the spherically symmetric solution, we can get $r_0=(3k_1/(2q\kappa\rho-2\lambda))^{1/3}$ with $k_1$, $q$ and $\lambda$ being parameters. In particular $q$ represents the corrections to the ``effective'' cosmological constant due to matter sources, $\lambda$ imitates the behavior of the cosmological constant and $k_1$ is related to the scale of the source-term. The ``effective'' cosmological constant can be taken from the combination $2q\kappa\rho-2\lambda$. Take into account that in $f(R)$-gravity theories, the effects of an ``effective'' cosmological constant can be achieved by higher-order derivative contributions \cite{suggested2, suggested4} 
There exists the possibility that any theory able to reproduce the dark-energy effects, contains the scale $r_0$ in terms of the parameters involved in the theory at the local physics level. This will be a matter of investigation in coming works. With the Vainshtein conditions formulated as extremal conditions, then the corresponding mechanism can be expressed in terms of the "gauge" functions $T_0(r,t)$. In resume the mechanism can be understood by the set of conditions (\ref{eq:setofcond}) obtained with the help of (\ref{eq:cond1}). 
The equivalence between the extremal condition of the dynamical metric and the corresponding condition for the massive action in dRGT is obtained in eq. (\ref{eq:ptm}). This equivalence just demonstrates that the Vainshtein scale can be obtained as an extremal condition for the massive action. However, when all the degrees of freedom are inside the dynamical metric, this is equivalent to the same conditions but applied to the dynamical metric. Finally, I have also derived the equations of motion for a massive test particle moving under the influence of the dynamical metric (with all the degrees of freedom) in dRGT. If the one-dimensional equation of motion for the massive test particle is written in terms of the extended notion of energy defined in eq. (\ref{eq:Ronaldo}), then at the background level no-difference will appear with respect to GR for the three-dimensional motion of a test particle. However, as soon as some dynamics is involved for the system, then the differences between GR and dRGT will be more evident. The dynamical processes can be Hawking radiation, perturbation theory, evolving solutions without symmetry under time-translations, etc. \\\\        

{\bf Acknowledgement}

The author would like to thank Gia Dvali for a very useful discussion during the Karl Schwarzschild meeting 2013 organized in FIAS, Frankfurt/Germany. This work is supported by MEXT (The Ministry of Education, Culture, Sports, Science and Technology) in Japan and KEK Theory Center.

\end{document}